# Ensemble Transfer Learning for the Prediction of Anti-Cancer Drug Response

Yitan Zhu[1*], Thomas Brettin[1], Yvonne A. Evrard[2], Alexander Partin[1], Fangfang Xia[1], Maulik Shukla[1], Hyunseung Yoo[1], James H. Doroshow[3], Rick Stevens[1,4]

1. Computing, Environment and Life Sciences, Argonne National Laboratory, Lemont, IL, USA

2. Frederick National Laboratory for Cancer Research, Leidos Biomedical Research, Inc. Frederick, MD, USA

3. Developmental Therapeutics Branch, National Cancer Institute, Frederick, MD, USA

4. Department of Computer Science, The University of Chicago, Chicago, IL, USA

[*]Correspondence: yitan.zhu@anl.gov

# Abstract

Transfer learning has been shown to be effective in many applications in which training data for the target problem are limited but data for a related (source) problem are abundant. In this paper, we apply transfer learning to the prediction of anti-cancer drug response. Previous transfer learning studies for drug response prediction focused on building models that predict the response of tumor cells to a specific drug treatment. We target the more challenging task of building general prediction models for transfer learning that can make predictions for both new tumor cells and new drugs. While existing works focused on either building transformations of features and prediction targets between datasets or combining the source dataset with some auxiliary dataset for prediction, we apply the classic transfer learning framework that trains a prediction model on the source dataset and refines it on the target dataset, and extends the framework through ensemble. We implement the ensemble transfer learning framework using LightGBM and two deep neural network (DNN) models with different architectures. Uniquely, we investigate the power of transfer learning for three application settings including drug repurposing, precision oncology, and new drug development, through different data partition schemes in cross-validation. We test the proposed ensemble transfer learning on benchmark in vitro drug screening datasets, taking one dataset as the source domain and another dataset as the target domain. The analysis results demonstrate the benefit of applying ensemble transfer learning for predicting anti-cancer drug response in all three applications with both LightGBM and DNN models. Compared between the different prediction models, a DNN model with two subnetworks for the inputs of tumor features and drug features separately outperforms LightGBM and the other DNN model that concatenates tumor features and drug features for input in the drug repurposing and precision oncology applications. In the more challenging application of new drug development, LightGBM performs better than the other two DNN models, probably due to the limited number of drugs in the training set.

# Introduction

Cancer is a complex, dynamic, and heterogenous disease. Patients with the same cancer histology can respond differently to the same anti-cancer therapy [Wu et al., 2017]. Multiple in vitro drug screening studies have been conducted generating data about drug efficacy on cancer cell lines (CCLs) [Shoemaker, 2006; Basu et al., 2013; Yang et al., 2013; Barretina et al., 2012; Haverty et al., 2016]. Due to the heterogeneity of cancer, an accurate prediction of the response of cancer cells to a drug treatment is of paramount importance for therapeutics development and patient care. There are three major applications for drug response prediction including drug repurposing, precision oncology, and new drug development. The goal of drug repurposing is to examine whether an existing drug used to treat a specific cancer indication can be used to treat another cancer indication. In drug repurposing, both the drug and cancer are not new but their combination has not been previously tested. For precision oncology, the goal is to identify an existing drug to treat a new cancer case that has not been investigated or treated before. The development of new drugs requires predicting the response of known cancer cases under the treatment of a new drug that has not been tested before.

Various methods and analysis schemes have been developed and used for predicting anti-cancer drug response, which can be categorized in different ways. Conventional machine learning methods, such as ridge and elastic net regressions [Jang et al., 2014], random forests regression [Costello et al., 2014], and support vector machine [Huang et al., 2017], have been used in drug response prediction. Recently, deep learning methods have started to play an increasingly important role [Xia et al., 2018; Manica et al., 2019; Rampášek et al., 2019; Chang et al., 2018]. Some studies predicted dose-dependent cell growth inhibition [Xia et al., 2018], and many others predicted dose-independent drug response measurements, such as the area under the dose response curve (AUC) and the half maximal inhibitory concentration (IC$_{50}$) [Rampášek et al., 2019; Menden et al., 2013; Huang et al., 2017]. Some analyses have constructed a prediction model for an individual cancer type and/or drug [Rampášek et al., 2019; Smith et al., 2010; Fowles et al., 2016], while others have built general prediction models covering multiple cancer types and/or drugs [Xia et al., 2018; Chang et al., 2018; Menden et al., 2013; Manica et al. 2019]. While transcriptomic data and other omics data, such as genome and proteomic data, have been used for the prediction of drug response, transcriptomic data have been shown to be the most predictive among all omic modalities [Costello et al., 2014; Jang et al., 2014]. Most works have targeted the prediction of single drug response [Lee et al., 2007; Menden et al., 2013; Rampášek et al., 2019], though some predicted the response of drug combinations [Xia et al., 2018; Menden et al., 2019].

In this paper, we investigate the application of transfer learning for drug response prediction. The general goal of transfer learning is to build a high-performance learner for a target domain where data availability is limited using information from a related source domain with abundant data [Weiss et al., 2016; Pan et al., 2010]. Transfer learning has been used in many areas, such as text classification [Duan et al., 2012; Wang et al., 2011] and image classification [Duan et al., 2012; Kulis et al., 2011]. Deep transfer learning implements transfer learning with deep neural network (DNN) models [Tan et al., 2018; Huang et al., 2013; Oquab et al., 2014]. One popular deep transfer learning technique is to transfer the front layers of a DDN model trained in the source domain to the target domain and use it as a feature extractor [Oquab et al., 2014; Huang et al., 2013]. Based on the target domain data, either the parameters of the back layers are refined or the back layers are removed and new layers are added behind the front layers and trained from scratch. The idea behind this approach is that the DNN model forms an iterative and continuous abstraction process and the front layers may generate features informative in both domains [Tan et al., 2018]. Transfer learning has also been used for drug response prediction. Dhruba et al. utilized one drug screening dataset to help the prediction on another drug screening dataset through transfer learning, which either transforms the two datasets into a unified latent space or transforms one dataset to the space of the other dataset through regression mappings [Dhruba et al., 2018]. Turki et al. developed approaches to combine a drug screening dataset with auxiliary data for predicting patient treatment response [Turki et al., 2017; Turki et al., 2018]. Borisov et al. predicted the response of a patient to a drug treatment by building a prediction model for the patient using cell lines similar to the patient evaluated by gene expressions of selected drug-related pathways [Borisov et al., 2018].

While existing works on transfer learning for drug response prediction focus on building prediction models for a specific drug [Dhruba et al., 2018; Turki et al., 2017; Turki et al., 2018; Borisov et al., 2018], we target the more challenging task of building general prediction models through transfer learning that can predict the response of not only new cancer cases but also new drugs. Uniquely, we test the power of transfer learning for three drug response prediction applications including drug repurposing, precision oncology, and new drug development, via different data partition schemes in cross-validation. Also, different from the previous studies, we apply the classic transfer learning scheme that trains a prediction model on the source dataset and then refines it on the target dataset, and extend the scheme through ensemble prediction by training and refining multiple models. We implement the analysis pipeline using three prediction models, including LightGBM [Ke et al., 2017], a representative and efficient gradient boosting algorithm, and two DNN models of different architectures. We apply ensemble transfer learning on multiple in vitro CLL drug screening datasets simulating the three different drug response prediction applications. Based on the results, we compare the prediction performance with and without transfer learning and also compare between transfer learning using different prediction models for each application.

## Data and Methods

### *Drug Response Data, Gene Expressions, and Drug Descriptors*

Our study involves four public in vitro drug screening datasets, including the Cancer Therapeutics Response Portal v2 (CTRP) [Basu et al., 2013], the Genomics of Drug Sensitivity in Cancer (GDSC) [Yang et al., 2013], the Cancer Cell Line Encyclopedia (CCLE) [Barretina et al., 2012], and the Genentech Cell Line Screening Initiative (GCSI) [Haverty et al., 2016]. The drug response values of these datasets are the percentages of tumor cell growth under a drug treatment at multiple doses. We used the three-parameter logistic function (hill slope model) to fit the tumor cell growth values and generate dose response curves. Based on the dose response curve, we calculated the area under the dose response curve (AUC) for the dose range of $[10^{-10}$ M, $10^{-4}$ M$]$. The AUC value was then normalized by the dose range, so that after normalization, the AUC value is between 0 and 1, representing the treatment effect. 0 indicates complete response and 1 indicates no response. Table 1 shows the number of treatments (pairs of drugs and CCLs) in each dataset.

In some studies, a drug and CCL pair may have been tested multiple times. In these cases, we averaged the AUC values across the experiments so that the number of treatments in Table 1 reflects the number of unique drug and CCL pairs in a dataset. Fig. 1 shows the histogram of AUC values with the mean and standard deviation in each dataset. Clearly, the distribution of AUC values varies between datasets.

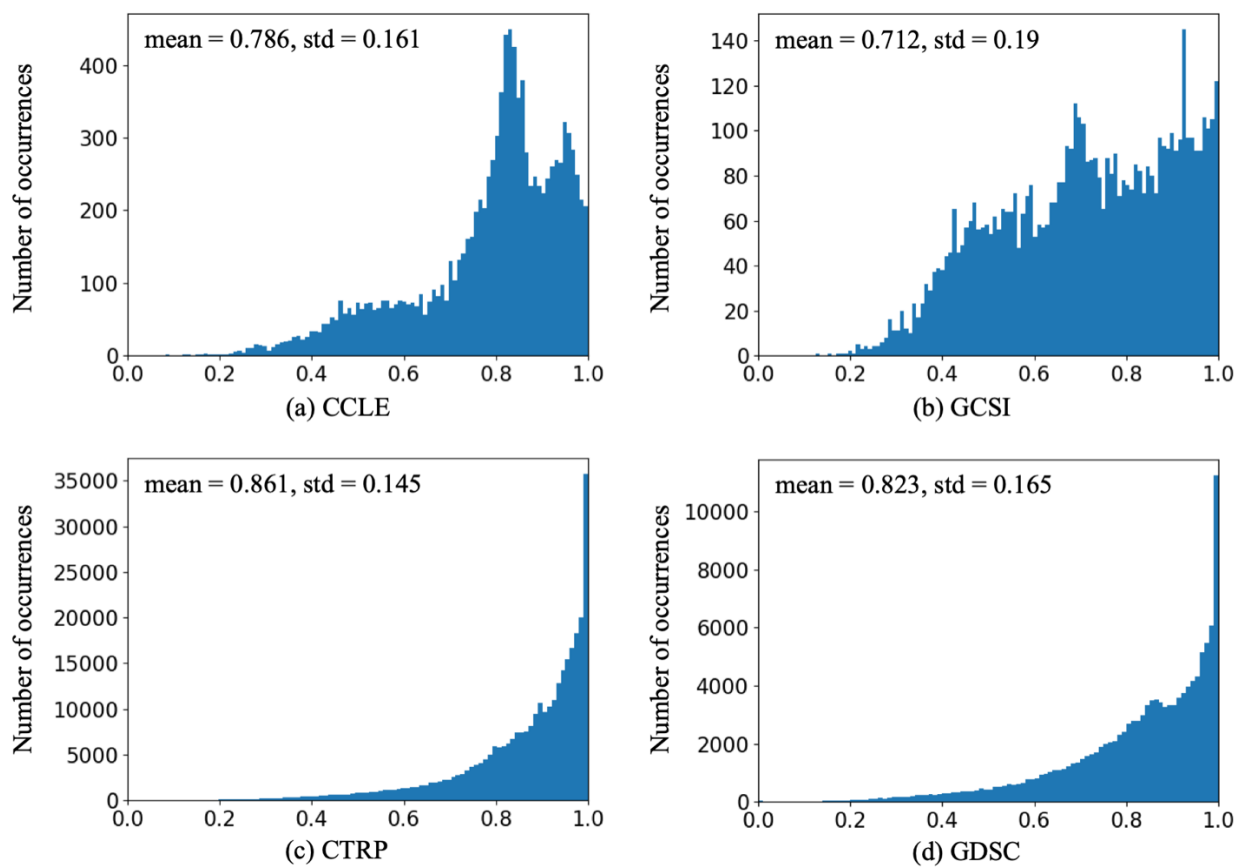

**Figure 1** Histograms of drug response AUC values in datasets. Mean and standard deviation (std) of AUC values are shown on the top left of each histogram.

CCLs are represented by their gene expression data in prediction modeling. The gene expression data were collected from the CCLE online resource. All CCLs in the other three studies, i.e., GCSI, CTRP, and GDSC, were also used in the CCLE study, except 11 GDSC CCLs that were thus excluded from the analysis. The gene expression data were generated using RNA sequencing, and TPMs (transcripts per kilobase million) were calculated as expression values, which were log2 transformed and then standardized so that each gene has a 0 mean and a unit standard deviation. Instead of using all transcripts for analysis, we focused the analysis on genes potentially related to cancer genetic mechanism, genomic regulation, and drug response. We selected 1,927 genes, including "landmark" genes well-representing cellular transcriptomic changes identified in the Library of Integrated Network-Based Cellular Signatures (LINCS) project [Subramanian et al., 2017] and cancer-related genes collected from OncoKB [Chakravarty et al., 2017] and GDSC [Iorio et al., 2016].

Drugs are represented by molecular descriptors in prediction modeling. The Dragon (version 7.0) software package (https://chm.kode-solutions.net/products_dragon.php) was used to compute numeric descriptors of the drugs based on their molecular structure. The package calculated various types of descriptors, such as the simplest atom types, functional groups and fragment counts, topological and geometrical descriptors, estimations of molecular properties, and drug-like and lead-like indices. We removed the descriptors with missing values and kept 1,623 molecular descriptors for the analysis.

**Table 1.** Numbers of CCLs, drugs, and treatments (pairs of drugs and CCLs) in each dataset.

| Dataset | # CCLs | # Drugs | # Treatments |
|---|---|---|---|
| GCSI | 357 | 16 | 5,647 |
| CCLE | 474 | 24 | 10,971 |
| GDSC | 659 | 238 | 125,712 |
| CTRP | 812 | 494 | 318,040 |

*Framework of Analysis Scenario*

A goal of our study is to investigate whether ensemble transfer learning can improve the prediction of drug response compared to not using transfer learning. The ensemble transfer learning (ETL) first train prediction models on the source dataset and then refine them on a part of the target dataset. After refinement, the models are applied on the rest of the target dataset to make ensemble predictions. The prediction performance is then evaluated and compared to those of baseline schemes that build prediction models based on only the target data. Two baseline schemes without transfer learning are used, standard cross-validation (SCV) and ensemble cross-validation (ECV). The analysis schemes of SCV and ECV will be introduced in detail later. The prediction performances of the three analysis schemes are compared to each other. See Fig. 2 for the framework of the analysis scenario. For a fair comparison, the data partition used for model training, validation, and testing in the baseline schemes are exactly the same as the data partition used for model refinement, validation, and testing in transfer learning on the target dataset in corresponding cross-validation trials, respectively. The validation set is used for hyperparameter tuning and early stopping of model training or refinement. In Fig. 2, 8-1-1 cross-validation means dividing the data into 10 data folds and using 8, 1, and 1 data fold for model training, validation, and testing, respectively. 8-1-1 cross-validation is used at the first step of transfer learning to train models on the source dataset. 1-1-8 cross-validation means dividing the data into 10 data folds and using 1, 1, and 8 data folds for model training (or refinement), validation, and testing, respectively. 1-1-8 cross-validation is used for all analyses on the target data, including SCV, ECV, and the second step of transfer learning, to simulate a situation where the training data at the target domain are quite limited. For transfer learning, we use the two large datasets CTRP and GDSC (see Table 1) as the source data and use the two small datasets CCLE and GCSI as the target data, which forms 4 transfer learning tasks denoted by CTRP → CCLE, CTRP → GCSI, GDSC → CCLE, and GDSC → GCSI.

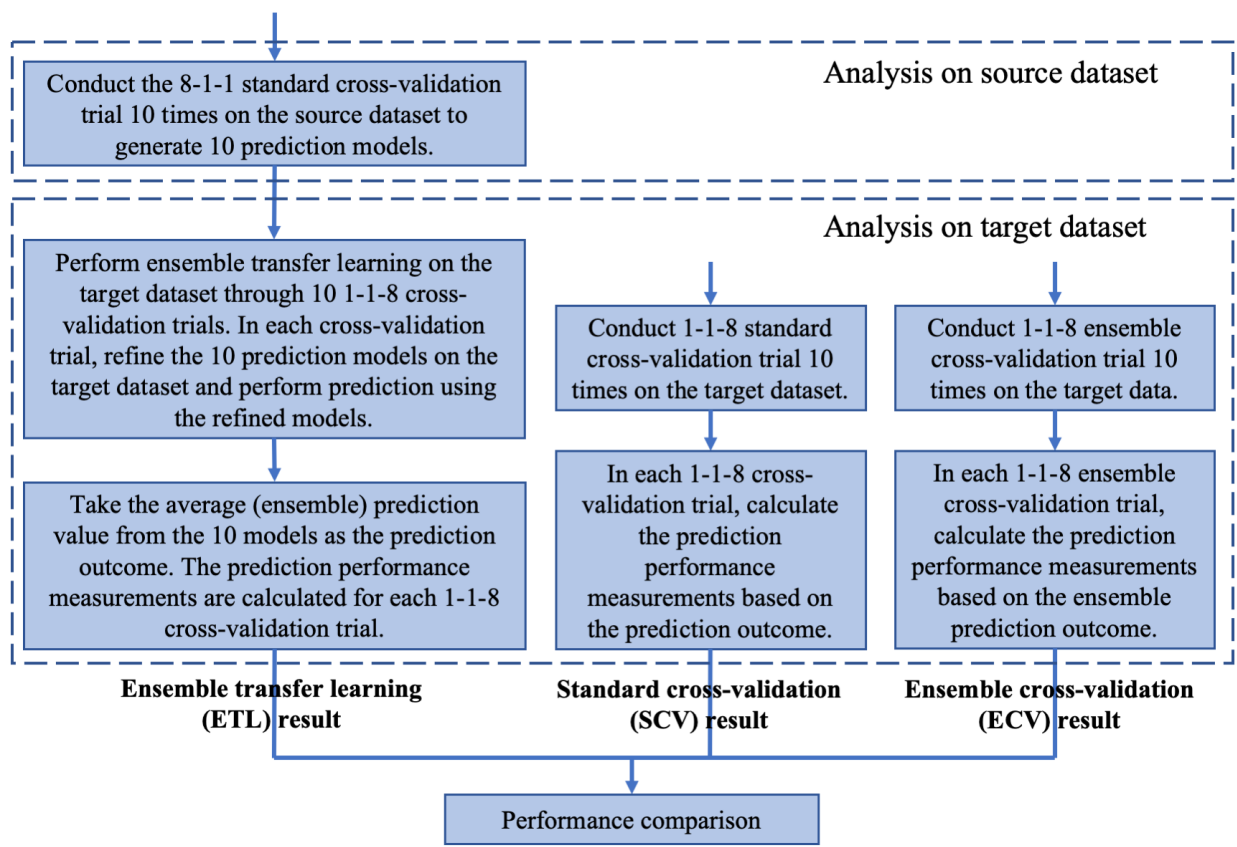

**Figure 2**   Analysis scenario framework. The analysis scheme on the left is ensemble transfer learning (ETL). The middle and right analysis schemes are standard cross-validation (SCV) and ensemble cross-validation (ECV), respectively, which do not apply transfer learning but instead analyze only the target dataset.

## *Standard and Ensemble Cross-Validations*

Fig. 3a shows the flowchart for standard cross-validation (SCV), in which data are divided into three parts for model training, validation, and testing. In the first step of transfer learning, we apply the 8-1-1 SCV on the source dataset to generate the models to be transferred. The 1-1-8 SCV is applied on the target data as a baseline to be compared with transfer learning. The ensemble cross-validation (ECV) also follows the flowchart in Fig. 3a but with the part indicated by the dashed-line box replaced by the flowchart in Fig. 3b, which performs ensemble learning by resampling the training set. We apply the 1-1-8 ECV on the target data as a second baseline to be compared with transfer learning. Notice that SCV, ECV, and ETL on the same target dataset always use the same data partition (i.e., training, validation, and testing sets) in corresponding cross-validation trials so that the prediction performances obtained by the analyses can be compared.

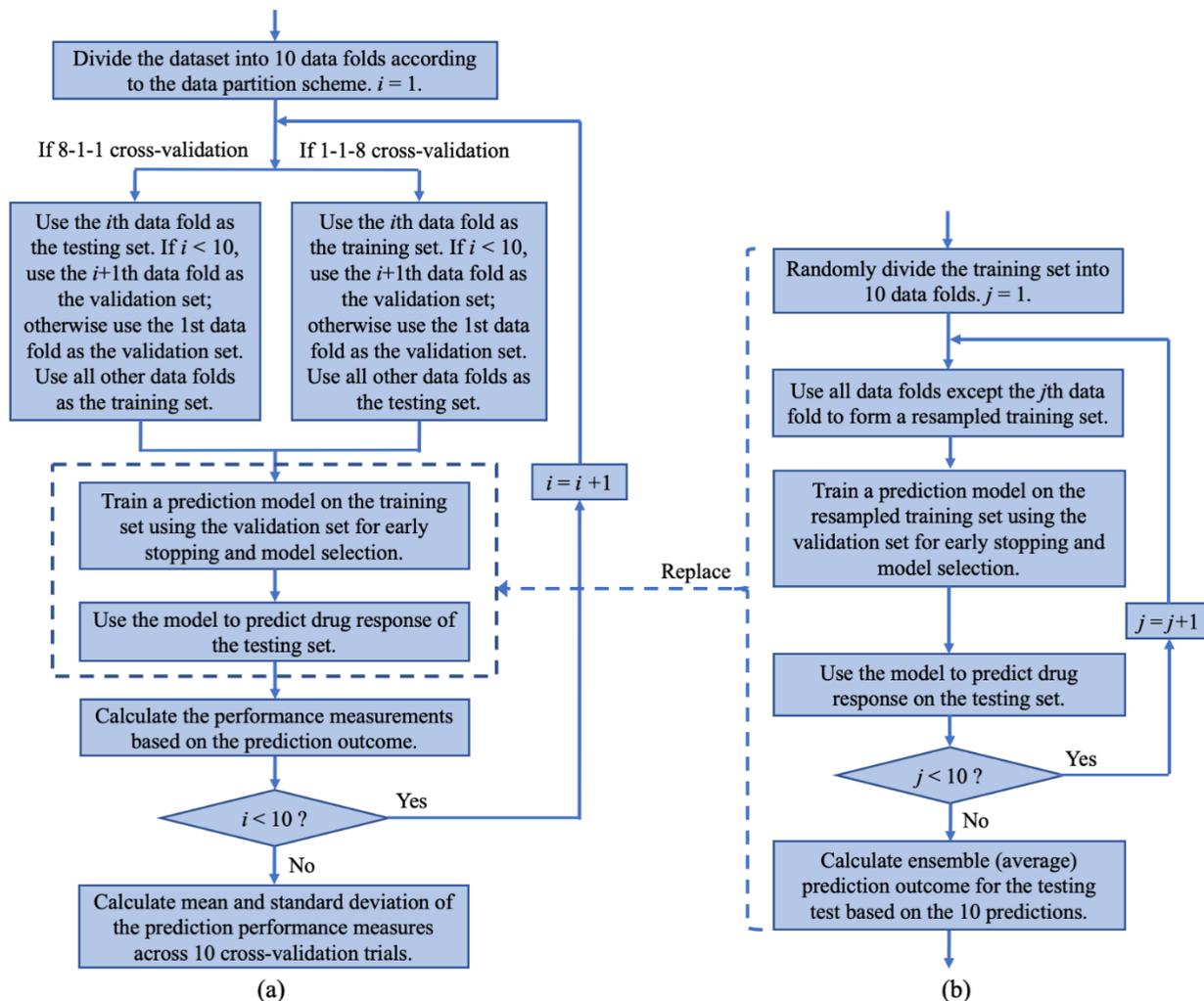

**Figure 3** (a) Flowchart of standard cross-validation (SCV). (b) The ensemble cross-validation (ECV) also follows the flowchart in (a), but with the part indicated by the dashed-line box replaced by the flowchart in (b).

## *Ensemble Transfer Learning Scheme*

Fig. 4 shows the flowchart for ensemble transfer learning (ETL), which retrieves the 10 models trained on the source dataset and refines these models on the training set of the target data. The refined models are then used to predict the testing samples in the target data, where their prediction outcomes are averaged to generate the ensemble prediction. We apply the ETL analysis for each of the four transfer learning tasks.

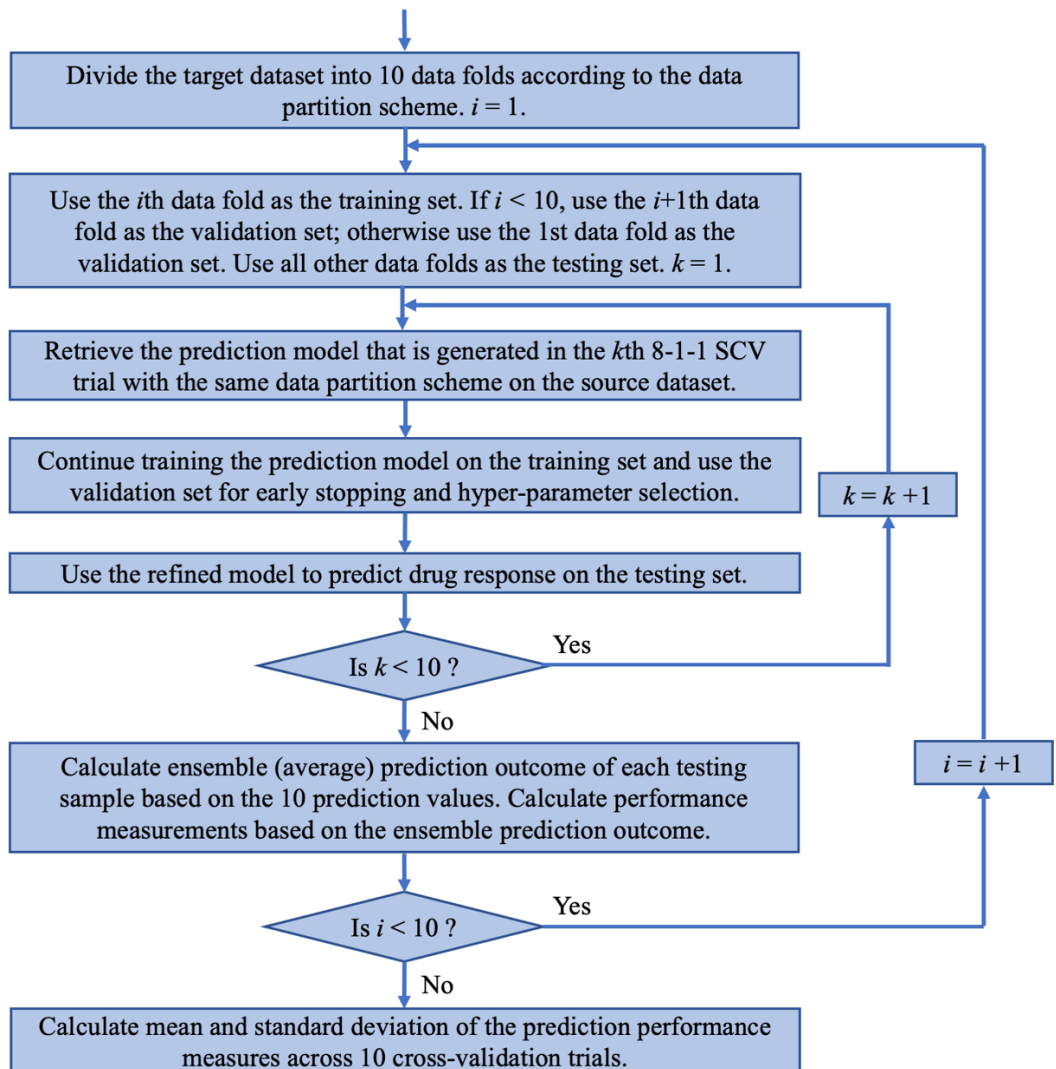

**Figure 4**   Flowchart of ensemble transfer learning (ETL).

## *Three Data Partition Schemes Representing Different Drug Response Prediction Applications*

We investigate the power of transfer learning for three different drug response prediction applications including drug repurposing, precision oncology, and new drug development. We design three data partition/selection schemes to simulate the three different applications for transfer learning tasks. For the purpose of evaluating generalization prediction performance, there should be no treatment (combination of CCL and drug) shared by the source and target datasets in analysis. Thus, we removed the overlapping treatments from the source dataset, so that they are included only in the target dataset. For drug repurposing, no additional data removal or selection was needed. See Section A of Table 2 for the numbers of CCLs, drugs, and treatments in the source dataset after removing overlapping treatments in each transfer learning task.

For the application of precision oncology, we further removed all treatments of CCLs from the source dataset that are also included in the target dataset, because the general goal of precision oncology is to select a drug for treating a tumor that has not been seen before. Also, when performing cross-validations on both the target and source datasets, the data folds were always generated to have random but different CCLs. In other words, no CCL was shared between data folds, which guaranteed that different CCLs were used for model training/refinement, validation, and testing, strictly simulating the precision oncology setup. See Section B of Table 2 for the numbers of CCLs, drugs, and treatments in the source dataset after removing overlapping CCLs in each transfer learning task. We can see the numbers of CCLs and treatments are significantly reduced compared to the numbers in Section A of Table 2.

For the application of new drug development, we removed all treatments of drugs from the source dataset that are also included in the target dataset, because the goal is to discover new drugs that can treat existing cancer cases. When performing cross-validations on both the target and source datasets, the data folds were always randomly generated to have different drugs, which guaranteed different drugs were used for model training/refinement, validation, and testing. See Section C of Table 2 for the numbers of CCLs, drugs, and treatments in the source dataset after removing overlapping drugs in each transfer learning task.

**Table 2.** Numbers of CCLs, drugs, and treatments in source datasets after removing overlap between source and target datasets, with different data partition/selection schemes.

| Transfer learning task | | Section A: removal of overlap treatments for drug repurposing | | | Section B: removal of overlap CCLs for precision oncology | | | Section C: removal of overlap drugs for new drug development | | |
|---|---|---|---|---|---|---|---|---|---|---|
| Target | Source | # CCLs | # Drugs | # Treatments | # CCLs | # Drugs | # Treatments | # CCLs | # Drugs | # Treatments |
| CCLE | CTRP | 812 | 494 | 311,194 | 376 | 494 | 143,634 | 812 | 477 | 305,278 |
| CCLE | GDSC | 659 | 238 | 123,447 | 282 | 238 | 53,743 | 659 | 225 | 121,174 |
| GCSI | CTRP | 812 | 494 | 314,469 | 479 | 494 | 185,138 | 812 | 482 | 309,363 |
| GCSI | GDSC | 659 | 238 | 123,141 | 343 | 238 | 64,664 | 659 | 224 | 120,045 |

## *DNN and LightGBM Prediction Models*

We take drug response prediction as a regression problem to predict the AUC value and use the mean squared error (MSE) as the loss function in training. LightGBM is an efficient implementation of the Gradient Boosting Decision Tree (GBDT) [Friedman, 2001] using techniques of gradient-based one-side sampling and exclusive feature bundling to speed up model training [Ke et al., 2017]. We used the LightGBM Python package (https://LightGBM.readthedocs.io/en/latest/index.html) for implementation. In transfer learning, the refinement of a LightGBM model was realized by adding additional boosting steps (decision trees) to fit the training set of the target data. The model training/refinement process would be stopped early if the loss on the validation set did not reduce in 150 boosting steps; otherwise the whole process took 1,500 boosting steps. For the other parameters of the LightGBM model, we used the default values.

We used the Keras package (https://keras.io/) with Tensorflow (https://www.tensorflow.org/) backend for implementing DNN models. Two DNN models with different architectures were implemented (see Fig. 5). The first DNN model is composed of 7 hidden fully connected (dense) layers with the number of nodes consecutively halved from the first hidden layer to the last hidden layer (Fig. 5a). The gene expressions of a CCL and the drug descriptors of a drug are concatenated to form the input. The second DNN model contains two subnetworks of 3 hidden dense layers, one for the input of gene expressions and the other for the input of drug descriptors (Fig. 5b). The outputs of the two subnetworks are concatenated and then passed to the other 4 hidden dense layers before output. The number of nodes is also consecutively halved from the first hidden layer to the last hidden layer. For convenience, we use sDNN (single-network DNN) and tDNN (two-subnetwork DNN) to denote the first and second DNN models, respectively. Both sDNN and tDNN have 7 hidden layers. Notice that although the total number of nodes in a hidden layer in tDNN is always larger than the number of nodes in the corresponding hidden layer in sDNN, the total number of trainable parameters in tDNN is significantly smaller than that of sDNN due to the subnetwork structure. In both models, each hidden layer has a dropout layer following it except the last hidden layer. All dropout layers in a model use the same dropout rate. In the analysis, the dropout rate was selected among 0, 0.1, 0.25, 0.45, and 0.7 by minimizing the validation loss. It was the only hyperparameter optimized in the model learning process. The Adam optimizer was used with default setting for model learning [Ba et al., 2015]. The learning rate was initialized at 0.001 and was reduced by a factor of 10 if the reduction of validation loss was smaller than 0.00001 in 10 epochs. The learning process would be early stopped if the reduction of validation loss was smaller than 0.00001 in 20 epochs; otherwise the full learning process would take 100 epochs. When refining a trained DNN model for transfer learning, we kept the parameters of the bottom 2 hidden layers unchanged and continued training the parameters associated with the top 5 hidden layers on the target dataset. The dropout rate was also re-selected among the five candidate values based on the validation loss.

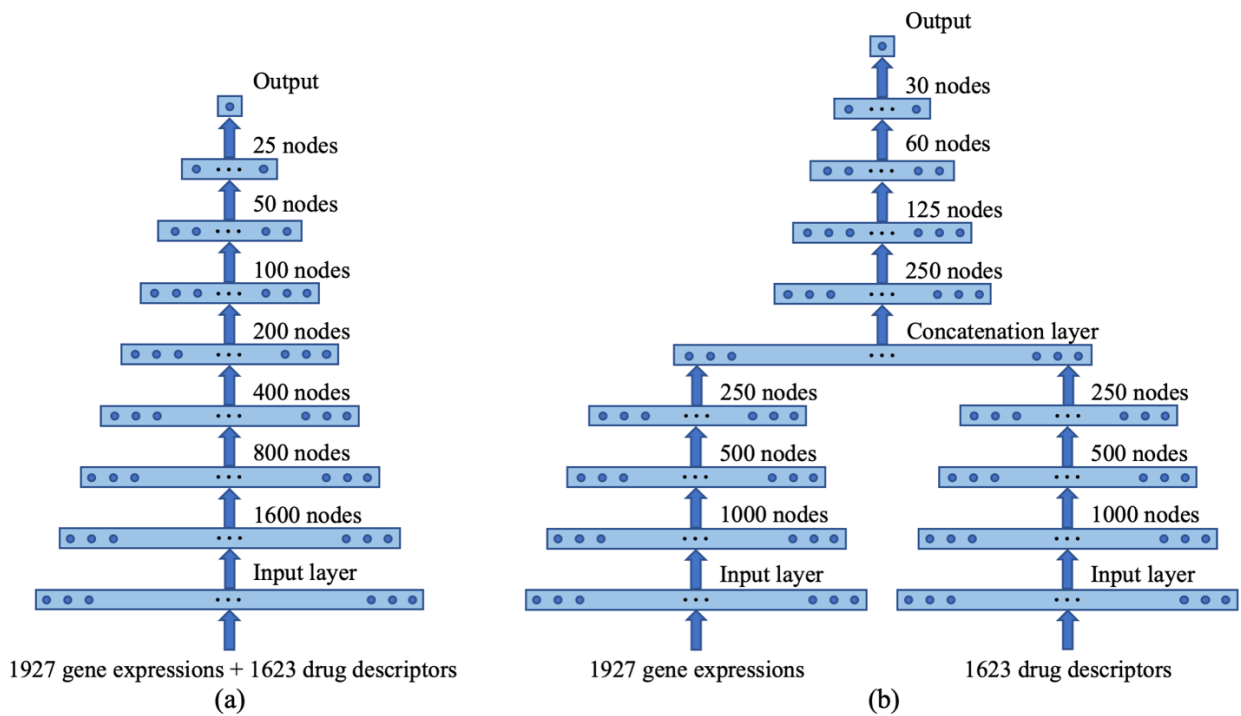

**Figure 5** Architectures of two DNN models used in the analysis. (a) Single-network DNN (sDNN) model. Gene expressions and drug descriptors are concatenated to form the input. (b) Two-subnetwork DNN (tDNN) model. The subnetworks take gene expressions and drug descriptors as inputs separately.

## Results

### *Prediction Performance for Drug Repurposing Application*

For the drug repurposing application, we performed ensemble transfer learning (ETL), standard cross-validation (SCV), and ensemble cross-validation (ECV) with three prediction models including LightGBM, sDNN (single-network DNN), and tDNN (two-subnetwork DNN). ETL was conducted for all four transfer learning tasks (i.e., CTRP → CCLE, CTRP → GCSI, GDSC → CCLE, and GDSC → GCSI), while SCV and ECV were conducted on the two target datasets, i.e., CCLE and GCSI. We used two measures to evaluate the testing prediction performance. The first measure is the root of mean squared error (RMSE), which is the square root of the loss function used by the prediction models. The second measure is the Pearson correlation coefficient that evaluates the variation consistency between prediction values and true values. The prediction performance was evaluated 10 times in the 10 cross-validation trails for each of ETL, SCV, and ECV. Then we examined the difference of prediction performance between ETL and SCV/ECV using the pair-wise t-test based on the 10 measurements of each analysis.

See Table 3 for the obtained prediction performance and comparison. In all of the four transfer learning tasks, tDNN always outperforms the other two prediction models, i.e., lightGBM and sDNN, evaluated by both smaller average RMSE and larger average correlation coefficient. Two-tail t-tests show ETL with any prediction model always statistically significantly outperforms SCV and ECV with the same prediction model (p-values $\leq 0.05$). This indicates the benefit of using ensemble transfer learning for anti-cancer drug response prediction. ECV always gives a better prediction performance than SCV when applying the same prediction model on the same target dataset, which shows the advantage of ensemble learning.

*Prediction Performance for Precision Oncology Application*

Table 4 shows the prediction performance and comparison for the precision oncology application, with cross-validations based on hard partitioning of CCLs. Again, tDNN always outperforms the other two prediction models, except only when being evaluated by the correlation coefficient in the CTRP $\rightarrow$ CCLE learning task. Two-tail t-tests show ETL with any prediction model always statistically significantly outperforms SCV and ECV with the same model (p-values $\leq 0.05$), except only in the comparison of ETL (with sDNN) and ECV (with sDNN) in the learning task of GDSC $\rightarrow$ CCLE. These indicate the benefit of using ensemble transfer learning for drug response prediction in precision oncology applications.

*Prediction Performance for New Drug Development Application*

Table 5 shows the prediction performance and comparison for the new drug development application with cross-validations based on hard partitioning of drugs. Predicting the efficacy of new drugs is generally a more challenging task than predicting the response of new CCLs. Also, because there are not many drugs tested in the CCLE and GCSI studies (see Table 1), the number of drugs used for training or refining a prediction model on these two target datasets is no larger than 3, which forms a very difficult prediction problem. It is not surprising to see that the prediction performance of ETL is worse for new drug development than for precision oncology and drug repurposing. But interestingly ETL also shows a higher improvement on the prediction performance for new drug development than for the other two applications, evaluated by the difference between ETL and ECV/SCV performance measurements.

Compared among three different prediction models, tDNN performs best in the transfer learning task of CTRP $\rightarrow$ CCLE, while LightGBM performs best in the other three transfer learning tasks. This is different from the cases of drug repurposing and precision oncology, where tDNN almost always outperforms LightGBM and sDNN. A possible reason is that the LightGBM model has a model complexity lower than those of DNN models, measured by the number of trainable parameters. Thus, it is more generalizable for prediction on new drugs, especially when the training data include very few drugs. With any of the three prediction models, ETL always statistically significantly outperforms SCV and ECV (p-values $\leq 0.05$), except only the comparison of ETL (with sDNN) and SCV (with sDNN) for the transfer learning tasks on the GCSI dataset when the prediction performance is evaluated by the correlation coefficient. The result shows the benefit of using ensemble transfer learning for new drug development.

*Prediction Performance of Transfer Learning Using Individual Model Without Ensemble*

Since we have performed ensemble transfer learning (ETL), it is straightforward to calculate the prediction performance of transfer learning using an individual model trained on the source data without ensemble prediction, which is called standard transfer learning (STL). Detail results of STL cannot be presented here due to the large number of models trained in the analysis, but we can summarize the major observations on the results. For all three prediction models, STL sometimes does not produce a prediction performance better than those of SCV and ECV in the drug repurposing and precision oncology applications. But ETL has been shown to dominantly outperform SCV and ECV for these two applications, which indicates the importance of using transfer learning and ensemble prediction simultaneously for drug response prediction. For the more challenging application of new drug development, we find STL almost always outperforms SCV and ECV, while ETL further improves the prediction performance.

# Discussion

Compared to existing works, our study is the first research attempt of its kind, which can be summarized from three aspects. First, we investigate whether transfer learning improves the performance of general response prediction models for multiple cancer types and drugs including new drugs not used in model training, whereas existing works focus on building drug-specific prediction models through transfer learning. Our prediction task is more challenging. Second, we studied the power of transfer learning in three drug response prediction applications including drug repurposing, precision oncology, and drug development based on different data partition and selection schemes in cross-validation, which to our knowledge has not been investigated before. Third, unlike previous transfer learning studies that emphasize building transformations of features and drug response values between datasets [Dhruba et al., 2018], we study the power of the classic transfer learning scheme that trains a prediction model on the source data and then refines it on the target data. We also extend the classic scheme via ensemble and show the benefit of performing ensemble transfer learning. Although the distribution of drug response varies between datasets (Fig. 1) and the same treatments (pairs of drugs and CCLs) might have quite different response values in different datasets [Dhruba et al., 2018], indicating the existence of variation between datasets, transfer learning by model refinement on the target dataset seems to overcome this gap to certain extent.

We applied transfer learning with three different prediction models, including LightGBM, sDNN (single-network DNN), and tDNN (two-subnetwork DNN). In transfer learning with DNN models, we also tried freezing the parameters of the bottom 4 hidden layers and adjusting only the parameters associated with the top 3 hidden layers and the dropout rate in the model refinement stage. The obtained prediction performance was worse than what we got when freezing only the bottom 2 hidden layers, indicating the importance of having sufficient layers trainable in model refinement for transfer learning.

**Table 3** Comparison on the prediction performance of standard cross-validation (SCV), ensemble cross-validation (ECV), and ensemble transfer learning (ETL) for drug repurposing application

| Target | Source | Model | RMSE (SCV) | RMSE (ECV) | RMSE (ETL) | P-value (RMSE, SCV vs. ETL) | P-value (RMSE, ECV vs. ETL) | Cor (SCV) | Cor (ECV) | Cor (ETL) | P-value (Cor, SCV vs. ETL) | P-value (Cor, ECV vs. ETL) |
|---|---|---|---|---|---|---|---|---|---|---|---|---|
| CCLE | CTRP | lightGBM | 0.0895(0.0007) | 0.0872(0.0009) | 0.0827(0.0007) | 2.30E-11 | 1.36E-08 | 0.8313(0.0029) | 0.8403(0.0037) | 0.8581(0.0023) | 4.30E-11 | 1.82E-08 |
| CCLE | CTRP | sDNN | 0.0895(0.0013) | 0.0863(0.0010) | 0.0812(0.0007) | 3.13E-08 | 4.17E-07 | 0.8341(0.0050) | 0.8466(0.0045) | 0.8672(0.0030) | 1.97E-08 | 4.72E-07 |
| CCLE | CTRP | tDNN | 0.0918(0.0009) | 0.0867(0.0009) | **0.0756**(0.0005) | 4.96E-12 | 7.66E-11 | 0.8236(0.0033) | 0.8435(0.0030) | **0.8841**(0.0025) | 2.85E-12 | 3.81E-11 |
| CCLE | GDSC | lightGBM | 0.0895(0.0007) | 0.0872(0.0009) | 0.0839(0.0009) | 5.06E-09 | 2.52E-07 | 0.8313(0.0029) | 0.8403(0.0037) | 0.8535(0.0035) | 8.89E-10 | 3.13E-08 |
| CCLE | GDSC | sDNN | 0.0895(0.0013) | 0.0863(0.0010) | 0.0838(0.0008) | 3.71E-07 | 2.43E-06 | 0.8341(0.0050) | 0.8466(0.0045) | 0.8562(0.0037) | 6.10E-07 | 1.42E-05 |
| CCLE | GDSC | tDNN | 0.0918(0.0009) | 0.0867(0.0009) | **0.0811**(0.0007) | 2.85E-10 | 9.30E-08 | 0.8236(0.0033) | 0.8435(0.0030) | **0.8654**(0.0022) | 1.25E-10 | 1.55E-08 |
| GCSI | CTRP | lightGBM | 0.1168(0.0005) | 0.1142(0.0007) | 0.1063(0.0015) | 2.08E-09 | 3.89E-08 | 0.7889(0.0018) | 0.7992(0.0017) | 0.8293(0.0048) | 2.11E-10 | 3.85E-09 |
| GCSI | CTRP | sDNN | 0.1167(0.0025) | 0.1119(0.0017) | 0.1051(0.0014) | 7.05E-07 | 1.57E-06 | 0.7956(0.0111) | 0.8118(0.0057) | 0.8384(0.0047) | 2.13E-06 | 1.26E-06 |
| GCSI | CTRP | tDNN | 0.1177(0.0032) | 0.1109(0.0014) | **0.0962**(0.0018) | 7.93E-09 | 4.92E-09 | 0.7923(0.0105) | 0.8133(0.0050) | **0.8633**(0.0055) | 1.72E-08 | 4.97E-09 |
| GCSI | GDSC | lightGBM | 0.1168(0.0005) | 0.1142(0.0007) | 0.1059(0.0015) | 1.99E-09 | 4.93E-08 | 0.7889(0.0018) | 0.7992(0.0017) | 0.8321(0.0056) | 7.83E-10 | 7.97E-09 |
| GCSI | GDSC | sDNN | 0.1167(0.0025) | 0.1119(0.0017) | 0.1047(0.0021) | 4.92E-06 | 1.57E-05 | 0.7956(0.0111) | 0.8118(0.0057) | 0.8419(0.0048) | 2.43E-06 | 9.13E-07 |
| GCSI | GDSC | tDNN | 0.1177(0.0032) | 0.1109(0.0014) | **0.0995**(0.0017) | 1.76E-09 | 1.12E-09 | 0.7923(0.0105) | 0.8133(0.0050) | **0.854**(0.0052) | 1.54E-09 | 9.77E-11 |

RMSE indicates the square root of mean square error. Cor indicates the Pearson correlation coefficient. In the RMSE and Cor columns, the number before a parenthesis is the average prediction performance and the number in a parenthesis is the standard deviation, calculated across 10 cross-validation trials. The p-values are generated by pairwise t-tests and indicate how significantly the prediction performance of ETL differs from those of SCV and ECV. SCV vs. ETL indicates comparison of SCV and ETL. ECV vs. ETL indicates comparison of ECV and ETL. The best average prediction performance for each transfer learning task is indicated with bold.

**Table 4** Comparison on the prediction performance of standard cross-validation (SCV), ensemble cross-validation (ECV), and ensemble transfer learning (ETL) for precision oncology application

| Target | Source | Model | RMSE (SCV) | RMSE (ECV) | RMSE (ETL) | P-value (RMSE, SCV vs. ETL) | P-value (RMSE, ECV vs. ETL) | Cor (SCV) | Cor (ECV) | Cor (ETL) | P-value (Cor, SCV vs. ETL) | P-value (Cor, ECV vs. ETL) |
|---|---|---|---|---|---|---|---|---|---|---|---|---|
| CCLE | CTRP | lightGBM | 0.0913(0.0015) | 0.0894(0.0015) | 0.087(0.0016) | 6.43E-06 | 1.08E-04 | 0.8245(0.0045) | 0.8325(0.0047) | 0.8419(0.0056) | 3.75E-06 | 8.75E-05 |
| | | sDNN | 0.0915(0.0014) | 0.0886(0.0009) | 0.0858(0.0014) | 5.99E-06 | 5.27E-07 | 0.8275(0.0052) | 0.8385(0.0041) | **0.8479**(0.0049) | 3.67E-06 | 2.90E-05 |
| | | tDNN | 0.0909(0.0013) | 0.0882(0.0009) | **0.0856**(0.0014) | 2.39E-06 | 3.89E-05 | 0.8293(0.0040) | 0.8386(0.0038) | 0.8476(0.0037) | 2.69E-06 | 1.46E-04 |
| CCLE | GDSC | lightGBM | 0.0913(0.0015) | 0.0894(0.0015) | 0.0877(0.0014) | 5.37E-07 | 1.58E-04 | 0.8245(0.0045) | 0.8325(0.0047) | 0.8389(0.0045) | 6.01E-07 | 1.80E-04 |
| | | sDNN | 0.0915(0.0014) | 0.0886(0.0009) | 0.0888(0.0013) | 3.27E-04 | 3.28E-01 | 0.8275(0.0052) | 0.8385(0.0041) | 0.8366(0.0038) | 1.26E-05 | 4.86E-03 |
| | | tDNN | 0.0909(0.0013) | 0.0882(0.0009) | **0.0869**(0.0012) | 2.55E-05 | 3.87E-03 | 0.8293(0.0040) | 0.8386(0.0038) | **0.8428**(0.0040) | 1.22E-04 | 1.07E-02 |
| GCSI | CTRP | lightGBM | 0.1186(0.0023) | 0.116(0.0026) | 0.1118(0.0029) | 7.89E-05 | 1.75E-03 | 0.783(0.0090) | 0.7929(0.0094) | 0.8087(0.0109) | 9.27E-05 | 1.69E-03 |
| | | sDNN | 0.123(0.0043) | 0.1218(0.0033) | 0.1118(0.0016) | 1.25E-05 | 9.55E-06 | 0.7798(0.0160) | 0.7938(0.0082) | 0.8119(0.0049) | 1.14E-04 | 5.43E-05 |
| | | tDNN | 0.1237(0.0043) | 0.118(0.0029) | **0.1085**(0.0012) | 3.36E-06 | 5.19E-06 | 0.7804(0.0084) | 0.7989(0.0083) | **0.8228**(0.0042) | 2.83E-07 | 2.44E-05 |
| GCSI | GDSC | lightGBM | 0.1186(0.0023) | 0.116(0.0026) | 0.1099(0.0020) | 8.89E-08 | 6.55E-06 | 0.783(0.0090) | 0.7929(0.0094) | 0.8162(0.0072) | 1.59E-07 | 7.10E-06 |
| | | sDNN | 0.123(0.0043) | 0.1218(0.0033) | 0.1106(0.0016) | 6.42E-06 | 3.48E-06 | 0.7798(0.0160) | 0.7938(0.0082) | 0.8156(0.0063) | 3.76E-05 | 6.37E-07 |
| | | tDNN | 0.1237(0.0043) | 0.118(0.0029) | **0.1076**(0.0015) | 2.34E-06 | 6.80E-07 | 0.7804(0.0084) | 0.7989(0.0083) | **0.8258**(0.0046) | 4.06E-08 | 2.86E-07 |

RMSE indicates the square root of mean square error. Cor indicates the Pearson correlation coefficient. In the RMSE and Cor columns, the number before a parenthesis is the average prediction performance and the number in a parenthesis is the standard deviation, calculated across 10 cross-validation trials. The p-values are generated by pairwise t-tests and indicate how significantly the prediction performance of ETL differs from those of SCV and ECV. SCV vs. ETL indicates comparison of SCV and ETL. ECV vs. ETL indicates comparison of ECV and ETL. The best average prediction performance for each transfer learning task is indicated with bold.

**Table 5** Comparison on the prediction performance of standard cross-validation (SCV), ensemble cross-validation (ECV), and ensemble transfer learning (ETL) for the application of new drug development.

| Target | Source | Model | RMSE (SCV) | RMSE (ECV) | RMSE (ETL) | P-value (RMSE, SCV vs. ETL) | P-value (RMSE, ECV vs. ETL) | Cor (SCV) | Cor (ECV) | Cor (ETL) | P-value (Cor, SCV vs. ETL) | P-value (Cor, ECV vs. ETL) |
|---|---|---|---|---|---|---|---|---|---|---|---|---|
| CCLE | CTRP | lightGBM | 0.1828(0.0249) | 0.1826(0.0249) | 0.1589(0.0125) | 2.32E-02 | 2.42E-02 | 0.0739(0.0781) | 0.0778(0.0815) | 0.3742(0.1490) | 1.52E-04 | 1.45E-04 |
| | | sDNN | 0.2132(0.0608) | 0.1964(0.0460) | 0.158(0.0152) | 3.20E-02 | 4.52E-02 | 0.0762(0.0803) | 0.0685(0.1638) | 0.4455(0.0965) | 3.77E-05 | 8.99E-04 |
| | | tDNN | 0.206(0.0637) | 0.205(0.0602) | **0.1553**(0.0176) | 4.98E-02 | 4.46E-02 | 0.0917(0.1589) | 0.0937(0.1446) | **0.4667**(0.1172) | 1.20E-04 | 6.99E-05 |
| CCLE | GDSC | lightGBM | 0.1828(0.0249) | 0.1826(0.0249) | **0.1283**(0.0053) | 9.11E-05 | 9.57E-05 | 0.0739(0.0781) | 0.0778(0.0815) | **0.6301**(0.0525) | 2.52E-09 | 2.80E-09 |
| | | sDNN | 0.2132(0.0608) | 0.1964(0.0460) | 0.146(0.0201) | 1.04E-02 | 8.23E-03 | 0.0762(0.0803) | 0.0685(0.1638) | 0.5717(0.0539) | 5.76E-08 | 8.65E-06 |
| | | tDNN | 0.206(0.0637) | 0.205(0.0602) | 0.1412(0.0214) | 2.28E-02 | 1.92E-02 | 0.0917(0.1589) | 0.0937(0.1446) | 0.6124(0.0638) | 1.09E-05 | 6.65E-06 |
| GCSI | CTRP | lightGBM | 0.2491(0.0402) | 0.249(0.0401) | **0.1975**(0.0197) | 3.03E-03 | 3.02E-03 | 0.163(0.0643) | 0.1707(0.0599) | **0.396**(0.0366) | 1.41E-05 | 1.54E-05 |
| | | sDNN | 0.2804(0.0584) | 0.3042(0.0693) | 0.2243(0.0346) | 1.54E-02 | 1.57E-02 | 0.0172(0.2006) | -0.2031(0.1689) | 0.2231(0.1606) | 8.78E-02 | 1.27E-03 |
| | | tDNN | 0.3043(0.0931) | 0.2988(0.0670) | 0.215(0.0348) | 1.97E-02 | 6.88E-03 | -0.1835(0.1726) | -0.1463(0.2150) | 0.3707(0.0751) | 8.90E-06 | 1.17E-04 |
| GCSI | GDSC | lightGBM | 0.2491(0.0402) | 0.249(0.0401) | **0.2075**(0.0268) | 8.21E-03 | 8.23E-03 | 0.163(0.0643) | 0.1707(0.0599) | **0.3878**(0.0513) | 3.11E-05 | 3.63E-05 |
| | | sDNN | 0.2804(0.0584) | 0.3042(0.0693) | 0.2255(0.0239) | 2.42E-02 | 8.93E-03 | 0.0172(0.2006) | -0.2031(0.1689) | 0.1092(0.2477) | 4.89E-01 | 2.36E-03 |
| | | tDNN | 0.3043(0.0931) | 0.2988(0.0670) | 0.2148(0.0295) | 2.13E-02 | 7.11E-03 | -0.1835(0.1726) | -0.1463(0.2150) | 0.3147(0.0983) | 5.37E-06 | 2.87E-04 |

RMSE indicates the square root of mean square error. Cor indicates the Pearson correlation coefficient. In the RMSE and Cor columns, the number before a parenthesis is the average prediction performance and the number in a parenthesis is the standard deviation, calculated across 10 cross-validation trials. The p-values are generated by pairwise t-tests and indicate how significantly the prediction performance of ETL differs from those of SCV and ECV. SCV vs. ETL indicates comparison of SCV and ETL. ECV vs. ETL indicates comparison of ECV and ETL. The best average prediction performance for each transfer learning task is indicated with bold.

Prediction of anti-cancer drug response is a challenging task due to multiple reasons. First, the size of the currently available drug screening data is small compared to the huge dimensionality of cancer molecular/genomic system and the huge space of drug chemical structure, which forms a small-sample-size problem with insufficient data. Second, the drug mechanism of action (MoA) usually involves a complex chain of biochemical reactions and genomic regulations that might not be fully measured by existing experimental techniques. Thus, current data, such as gene expression and drug descriptors, might not be sufficient for modeling the drug action on tumors. Third, there is a gap between the current biological models used for drug screening and the therapeutic targets, patient tumors. The ultimate goal of predicting drug response is to either recommend an existing drug or design a new drug for treating a particular patient. Biological models currently used for drug screening mainly include CLLs and patient derived models, such as xenografts (PDXs) [Gao et al., 2015] and organoids (PDOrgs) [Aboulkheyr et al., 2018]. These biological models are different from each other and also different from the real patient tumors, leading to the variations of their drug responses. CLL drug screening data are relatively more abundant than both PDX/PDOrg drug screening data and patient treatment response data. Our work is a pilot study that builds general drug response prediction models through transfer learning, which can be helpful for future studies on transfer learning between different biological models and patients.

In future research, there are ways to potentially improve the drug response prediction accuracy. To better characterize the complex interaction between drug and tumor molecular system, existing knowledge of drug MoA and tumor genomic regulation mechanism can be integrated into the prediction model. Examples of drug MoA information are drug target genes and drug MoA categories. Examples of information about genomic regulation mechanism can be genetic pathways and protein-protein interactions affected by the drug MoA. Measurements or predictions on the binding affinities between drugs and target proteins may contribute to the prediction of drug response. Suitable feature selection and data representation may also be helpful. Gene selection methods such as co-expression extrapolation (COXEN) [Lee et al., 2007] can be used to identify genes that are both predictive for drug response and generalizable between different biological models and patient tumors. Chemical structures of drugs can be represented by not only numeric descriptors or fingerprints, but also by 2D/3D graphs/images that can be learned by deep neural networks.

## Conclusion

We developed the first ensemble transfer learning framework that builds general prediction models for predicting anti-cancer drug response. The transfer learning pipeline was implemented with three different prediction models including LightGBM, sDNN (single-network DNN), and tDNN (two-subnetwork DNN). Uniquely, we investigated the performance of the transfer learning pipeline for three drug response prediction applications including drug repurposing, precision oncology, and new drug development, based on in vitro drug screening data. Our results demonstrate the benefit of applying ensemble transfer learning in all of the three applications. For the comparison between three different prediction models, tDNN performs best in the drug repurposing and precision oncology applications, while LightGBM outperforms tDNN in 3 out of the 4 transfer learning tasks for the more challenging application of new drug development. Our work is a pilot study of transfer learning for building general drug response prediction models that are not specific to a particular drug, which may provide guidance for future research on transfer learning of drug response prediction between different biological models and patients.

## Funding

This work has been supported in part by the Joint Design of Advanced Computing Solutions for Cancer (JDACS4C) program established by the U.S. Department of Energy (DOE) and the National Cancer Institute (NCI) of the National Institutes of Health. This work was performed under the auspices of the U.S. Department of Energy by Argonne National Laboratory under Contract DE-AC02-06-CH11357, Lawrence Livermore National Laboratory under Contract DE-AC52-07NA27344, Los Alamos National Laboratory under Contract DE-AC5206NA25396, and Oak Ridge National Laboratory under Contract DE-AC05-00OR22725. This project has also been funded in whole or in part with federal funds from the National Cancer Institute, National Institutes of Health, under Contract No. HHSN261200800001E. The content of this publication does not necessarily reflect the views or policies of the Department of Health and Human Services, nor does mention of trade names, commercial products, or organizations imply endorsement by the U.S. Government.

## References

Aboulkheyr Es H, Montazeri L, Aref AR, et al. (2018) Personalized Cancer Medicine: An Organoid Approach. *Trends Biotechnol*. 2018 Apr;36(4):358-371. doi: 10.1016/j.tibtech.2017.12.005.

Ba J and Kingma D (2015) Adam: A method for stochastic optimization. In International Conference on Learning Representations (ICLR).

Barretina J, Caponigro G, Stransky N, et al. (2012) The Cancer Cell Line Encyclopedia enables predictive modelling of anticancer drug sensitivity. *Nature,* 2012 Mar 28; 483(7391):603-7. doi: 10.1038/nature11003.

Basu A, Bodycombe NE, Cheah JH, et al. (2013) An interactive resource to identify cancer genetic and lineage dependencies targeted by small molecules. *Cell*, 2013 Aug 29; 154(5):1151-1161. doi: 10.1016/j.cell.2013.08.003.

Borisov N, Tkachev V, Suntsova M, et al. (2018) A method of gene expression data transfer from cell lines to cancer patients for machine-learning prediction of drug efficiency. *Cell Cycle*. 2018;17(4):486-491. doi: 10.1080/15384101.2017.1417706.

Chakravarty D, Gao J, Phillips SM et al. (2017) OncoKB: A Precision Oncology Knowledge Base. *JCO Precis Oncol*. 2017 Jul;2017. doi: 10.1200/PO.17.00011. Epub 2017 May 16.

Chang Y, Park H, Yang HJ, et al. (2018) Cancer drug response profile scan (CDRscan): a deep learning model that predicts drug effectiveness from cancer genomic signature, *Scientific Reports*, volume 8, Article number: 8857

Costello JC, Heiser LM, Georgii E et al. (2014) A community effort to assess and improve drug sensitivity prediction algorithms. *Nat. Biotechnol.*, 32, 1202–1212.

Dhruba SR, Rahman R, Matlock K, et al. (2018) Application of transfer learning for cancer drug sensitivity prediction. *BMC Bioinformatics*. 2018 Dec 28; 19 (Suppl 17): 497. doi: 10.1186/s12859-018-2465-y.

Duan L, Xu D, Tsang IW. (2012) Learning with augmented features for heterogeneous domain adaptation. *IEEE Trans Pattern Anal Mach Intell*. 2012;36(6):1134–48.

Fowles JS, Brown KC, Hess AM, et al. (2016) Intra- and interspecies gene expression models for predicting drug response in canine osteosarcoma. *BMC Bioinformatics*. 2016 Feb 19;17:93. doi: 10.1186/s12859-016-0942-8.

Friedman JH. (2001) Greedy function approximation: a gradient boosting machine. *Annals of Statistics*, pages 1189–1232.

Gao H, Korn JM, Ferretti S, et al. (2015) High-throughput screening using patient-derived tumor xenografts to predict clinical trial drug response. *Nat Med*. 2015 Nov; 21(11):1318-25. doi: 10.1038/nm.3954.

Haverty PM, Lin E, Tan J, et al. (2016) Reproducible pharmacogenomic profiling of cancer cell line panels. *Nature*. 2016 May 19; 533(7603):333-7. doi: 10.1038/nature17987.

Huang JT, Li J, Yu D, et al. (2013) Cross-language knowledge transfer using multilingual deep neural network with shared hidden layers. In: Acoustics, Speech and Signal Processing (ICASSP), 2013 IEEE International Conference on. pp. 7304

Huang C, Mezencev R, McDonald JF et al. (2017) Open source machine-learning algorithms for the prediction of optimal cancer drug therapies. *PLoS One*. 2017 Oct 26;12(10):e0186906. doi: 10.1371/journal.pone.0186906.

Iorio F, Knijnenburg TA, Vis DJ, et al. (2016) A landscape of pharmacogenomic interactions in cancer. *Cell*. 2016 Jul 28;166(3):740-754. doi: 10.1016/j.cell.2016.06.017.

Jang IS, Neto EC, Guinney J et al. (2014) Systematic assessment of analytical methods for drug sensitivity prediction from cancer cell line data. *Pac. Symp. Biocomput*., 19, 63–74.

Ke G, Meng Q, Finley T, et al. (2017) LightGBM: A highly efficient gradient boosting decision tree, Advances in Neural Information Processing Systems, 3149–3157.

Kulis B, Saenko K, Darrell T. (2011) What you saw is not what you get: domain adaptation using asymmetric kernel transforms. In: IEEE 2011 conference on computer vision and pattern recognition. 2011. p. 1785–92.

Lee JK, Havaleshko DM, Cho H, et al. (2007) A strategy for predicting the chemosensitivity of human cancers and its application to drug discovery. *Proc Natl Acad Sci USA*, 2007 Aug 7; 104(32):13086-91. Epub 2007 Jul 31.

Manica M, Oskooei A, Born J, et al. (2019) Towards explainable anticancer compound sensitivity prediction via multimodal attention-based convolutional encoders, arXiv:1904.11223v3

Menden MP, Iorio F, Garnett M, et al. (2013) Machine learning prediction of cancer cell sensitivity to drugs based on genomic and chemical properties. *PLoS One*. 2013 Apr 30;8(4):e61318. doi: 10.1371/journal.pone.0061318.

Menden MP, Wang D, Mason MJ, et al. (2019) Community assessment to advance computational prediction of cancer drug combinations in a pharmacogenomic screen. *Nat Commun*. 2019 Jun 17;10(1):2674. doi: 10.1038/s41467-019-09799-2.

Oquab M, Bottou L, Laptev I, et al. (2014) Learning and transferring mid-level image representations using convolutional neural networks. In: Computer Vision and Pattern Recognition (CVPR), 2014 IEEE Conference on. pp. 1717

Pan SJ, Yang Q. (2010) A survey on transfer learning. *IEEE Trans Knowl Data Eng*. 2010;22(10):1345–59

Rampášek L, Hidru D, Smirnov P, et al. (2019) Dr.VAE: Improving drug response prediction via modeling of drug perturbation effects. *Bioinformatics*. 2019 Mar 8. pii: btz158. doi: 10.1093/bioinformatics/btz158.

Shoemaker RH (2006), The NCI60 human tumor cell line anticancer drug screen. *Nature Rev Cancer*, 2006; 6: 813-23.

Smith SC, Baras AS, Lee JK, et al. (2010) The COXEN principle: translating signatures of in vitro chemosensitivity into tools for clinical outcome prediction and drug discovery in cancer. *Cancer Res*. 2010 Mar 1;70(5):1753-8. doi: 10.1158/0008-5472.CAN-09-3562.

Subramanian A, Narayan R, Corsello SM, et al. (2017) A Next Generation Connectivity Map: L1000 Platform and the First 1,000,000 Profiles. *Cell*. 2017 Nov 30;171(6):1437-1452.e17. doi: 10.1016/j.cell.2017.10.049.

Tan C, Sun F, Kong T, et al. (2018) A survey on deep transfer learning. In International Conference on Artificial Neural Networks, pages 270–279. Springer, 2018.

Turki T, Wei Z, Wang JTL (2017) Transfer Learning Approaches to Improve Drug Sensitivity Prediction in Multiple Myeloma Patients. *IEEE Access* 5: 7381-7393, 2017. Doi: 10.1109/ACCESS.2017.2696523.

Turki T, Wei Z, Wang JTL (2018) A transfer learning approach via procrustes analysis and mean shift for cancer drug sensitivity prediction. *J Bioinform Comput Biol*. 2018 Jun;16(3):1840014. doi: 10.1142/S0219720018400140.

Wang C, Mahadevan S (2011) Heterogeneous domain adaptation using manifold alignment. In: Proceedings of the 22nd international joint conference on artificial intelligence, vol. 2. 2011. p. 541–46.

Weiss K, Khoshgoftaar TM, Wang D (2016) A survey of transfer learning. *Journal of Big Data*. volume 3, Article number: 9 (2016)

Wu D, Wang DC, Cheng Y, et al. (2017) Roles of tumor heterogeneity in the development of drug resistance: A call for precision therapy. *Semin Cancer Biol*. 2017 Feb; 42: 13-19. doi: 10.1016/j.semcancer.2016.11.006

Xia F, Shukla M, Brettin T, et al. (2018) Predicting tumor cell line response to drug pairs with deep learning. *BMC Bioinformatics*. 2018 Dec 21; 19 (Suppl 18): 486. doi: 10.1186/s12859-018-2509-3.

Yang W, Soares J, Greninger P, et al. (2013) Genomics of Drug Sensitivity in Cancer (GDSC): a resource for therapeutic biomarker discovery in cancer cells. *Nucleic Acids Res*, 2013 Jan; 41(Database issue):D955-61. doi: 10.1093/nar/gks1111